\documentclass[twocolumn]{aastex62}
\usepackage{graphicx}
\usepackage{color}
\usepackage{amsmath}
\usepackage{comment}
\usepackage{lineno}
\usepackage[first=0,last=9]{lcg}

\usepackage{array,multirow}

\definecolor{LightCyan}{rgb}{0.88,1,1}

\usepackage{appendix}
\usepackage{tabularx}
\usepackage{longtable}
\usepackage{enumerate}
\usepackage{enumitem}

\begin{document}

\title{\bf{Ancestral Black Holes of Binary Merger GW190521}}


\author{O. Barrera}
\affiliation{Department of Physics, University of Florida, PO Box 118440, Gainesville, FL 32611-8440, USA}


\author{I. Bartos}
\email{imrebartos@ufl.edu}
\affiliation{Department of Physics, University of Florida, PO Box 118440, Gainesville, FL 32611-8440, USA}

\begin{abstract}
GW190521 was the most massive black hole merger discovered by LIGO/Virgo so far, with masses in tension with stellar evolution models. A possible explanation of such heavy black holes is that they themselves are the remnants of previous mergers of lighter black holes. Here we estimate the masses of the ancestral black holes of GW190521, assuming it is the end product of previous mergers. We find that the heaviest parental black holes has a mass of $62^{+21}_{-19}$M$_\odot$ (90\% credible level). We find 78\% probability that it is in the $50$\,M$_\odot-120$\,M$_\odot$ mass gap, indicating that it may also be the end product of a previous merger. We therefore also compute the expected mass distributions of the "grandparent" black holes of GW190521, assuming they existed. Ancestral black hole masses could represent an additional puzzle piece in identifying the origin of LIGO/Virgo/KAGRA's heaviest black holes.
\end{abstract} 

\keywords{black holes, gravitational waves}

\section{Introduction} \label{sec:intro}

Some massive stars end their lives giving birth to black holes through stellar core collapse. Stars can produce black holes within a broad mass range, with a lower limit of about 5\,M$_\odot$. However, nuclear processes in some of the most massive stars are expected to lead to early stellar explosion, leaving no stellar remnants behind. This process can lead to a mass range of $\sim50$\,M$_\odot-120$\,M$_\odot$ with no remnant left behind \citep{2017ApJ...836..244W}. Nonetheless, the boundaries of this so-called pair-instability {\it mass gap} are currently uncertain \citep{2018ApJS..237...13L,2020ApJ...902L..36F,2020ApJ...905L..15B,2021MNRAS.501.4514C,2021arXiv210810885B}.

The more massive component of the binary merger GW190521 has the highest estimated mass ($95.3^{+28.7}_{-18.9}$\,M$_\odot$ assuming uninformative priors; \citealt{2020arXiv201014527A}) among the black holes detected by the LIGO \citep{2015CQGra..32g4001L} and Virgo \citep{2015CQGra..32b4001A} gravitational wave detectors. A similarly high mass is expected even if the binary's possible eccentricity is taken into account \citep{2020arXiv200905461G,2021arXiv210605575G,2020ApJ...903L...5R}. This black hole's mass is likely within the fiducial mass gap of $\sim50$\,M$_\odot-120$\,M$_\odot$, indicating that it might not have been formed directly through stellar evolution.

An attractive explanation for the high black hole mass is that it is itself the remnant of a previous merger of two, less massive black holes \citep{2017ApJ...840L..24F,2017PhRvD..95l4046G}. Such hierarchical mergers can occur in environments with large black hole number density such as galactic nuclei \citep{2020ApJ...893...35D,2020ApJ...900..177K,2021MNRAS.502.2049L}, and may be particularly common in active galactic nuclei (AGNs) that facilitate the further increase of the black hole density in the AGN accretion disk \citep{2012MNRAS.425..460M,2019PhRvL.123r1101Y,2020ApJ...890L..20G,2017ApJ...835..165B,2017MNRAS.464..946S}. The consecutive mergers of multiple black holes could explain the observed high masses even if these masses are inconsistent with stellar evolution. 

In this paper we derive the mass probability densities of the black holes that may have previously merged to produce the components of GW190521, assuming that GW190521 is indeed the latest stage of a chain of hierarchical mergers. We investigate the properties of the "parent" black holes and consider the possibility that GW190521 is a third generation merger, computing the expected properties of the "grandparent" black holes. While we focus on mass here, parental properties based on the black holes' spin was separately investigated by \cite{2021PhRvD.104h4002B}.
\newpage

\section{Computing the parental mass distribution}

We start with the posterior probability density $p(M)$ of a black hole mass $M$. Mass $M$ can be, for example, one of the component masses of GW190521. This posterior density depends on the reconstructed likelihood distribution $\mathcal{L}(M)$ based on the observed data and a prior distribution $\pi(M)$, such that $p(M)=\mathcal{L}(M)\pi(M)$. 

We neglected black hole spin here whose role we discuss separately below. Given $p(M)$ and assuming that $M$ is the merger remnant of two black holes with masses $m_1$ and $m_2\leq m_1$, we want to know the probability density $p(m_1,m_2|p(M))$. 

In the following we consider probability densities as a series of discrete values, reflecting the fact that they are obtained numerically. The probability that the parental masses fall into bin ${i,j}$ centered around $m_{1i},m_{2j}$ can be written as
\begin{equation}
p(m_{1i},m_{2j}|p(M)) = \sum_k p(m_{1i},m_{2j}|M_k)p(M_k),
\label{eq:1}
\end{equation}
where $p(M_k)$ is the probability of $M$ being in bin $k$ centered around mass $M_k$. We use Bayes' theorem to express the first term in the above sum as (see also \citealt{2021ApJ...914L..18D})
\begin{equation}
p(m_{1i},m_{2j}|M_k) = p(M_k|m_{1i},m_{2j})\frac{\pi(m_{1i},m_{2j})}{\pi(M_k)}
\label{eq:2}
\end{equation}
where $\pi(m_{1},m_{2})$ is the prior probability density of $\{m_1,m_2\}$ and 
\begin{equation}
\pi(M_k)=\sum_q \sum_r p(M_k|m_{1q},m_{2r})\pi(m_{1q},m_{2r})
\end{equation}
is the prior probability density of $M_k$. 

The probability $p(M_k|m_{1i},m_{2j})$ is 1 if the remnant of a binary {$m_{1i},m_{2j}$} has a mass in the $k$ bin of $M$, and 0 otherwise. Substituting Eq. \ref{eq:2} into \ref{eq:1} and marginalizing over $m_{2j}$ we obtain the posterior probability of $m_{1i}$:
\begin{equation}
p(m_{1i}|p(M))=\sum_j\sum_k p(M_k|m_{1i},m_{2j}) \pi(m_{1i},m_{2j})\frac{p(M_k)}{\pi(M_k)}.
\label{eq:3}
\end{equation}
A similar expression applies to $p(m_{2j}|p(M))$.

\begin{figure*}
\centering
 \includegraphics[angle = 0, scale = 0.54]{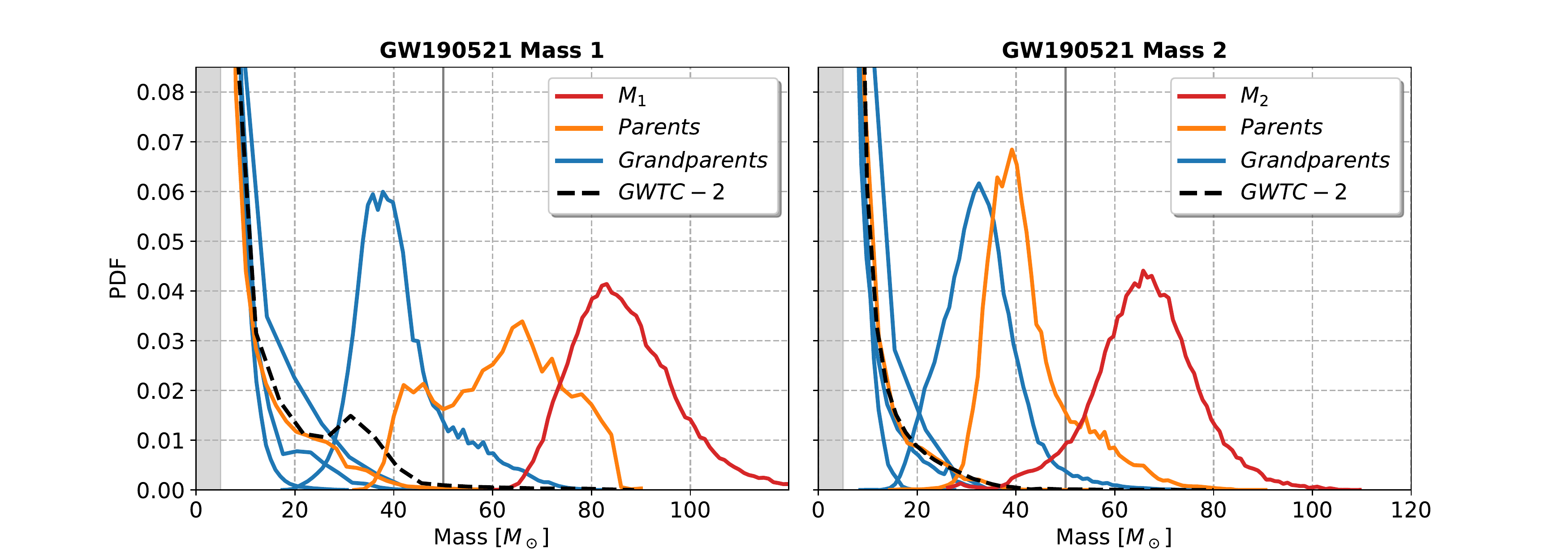}
      \caption{{\bf Probability density of ancestral black hole masses of GW190521,} separately for the more massive (left) and less massive (right) black hole in the binary. We show masses for two previous generations of mergers, and for comparison the mass distribution inferred from the GWTC-2 gravitational wave catalog (see legend).
    \label{fig:massPDF}}
\end{figure*}

\section{Remnant mass computation}

If we know the masses ($m_1,m_2$) and spins ($\vec{S}_1,\vec{S}_2$) of two black holes in a binary, the mass and spin of their remnant black hole can be determined. For this computation we use the phenomenological formulas of \cite{2014PhRvD..90j4004H}, who developed an analytical prescription for computing the remnant mass and spin based on a suite of numerical relativity simulations. 

Generally, both mass and spin are relevant in determining the properties of the remnant black hole. However, in the case of GW190521, the black hole spins are poorly determined, presenting only marginal constraints on the parental properties. Further, spin magnitudes and directions affect the mass loss by the binary due to gravitational wave emission by $\lesssim5\%$, i.e. much less than typical mass uncertainties. For grandparents and further generations, spin will be even more uncertain, therefore we neglect it here. \cite{2021PhRvD.104h4002B} considered the spin of GW190521 to constrain its parental black holes, and found that it is more consistent with at least one of the parents being itself the remnant of a merger. 

For a binary with well constrained black hole spins it is beneficial to extend our treatment to include spins as well \citep{2021PhRvD.104h4002B}. In particular, the ancestral mass ratio can have significant effect on the remnant black hole's spin \cite{2021PhRvD.104h4002B}.

Considering only the masses in reconstructing the properties of ancestral black holes for GW190521 gives us the function $p(M_k|m_{1i},m_{2j})$ used above. 

The phenomenological formulas of \cite{2014PhRvD..90j4004H} are accurate only if the mass ratio $m_2/m_1$ is not too small. We considered these formulas only for cases in which $m_2/m_1>0.33$. For more extreme mass ratios, we consider the mass loss by the binary through gravitational waves to be negligible, and adopt a remnant mass $M=m_1+m_2$. The mass loss in these cases is expected to be $<1\%$ which is much smaller than reconstruction uncertainties of the masses.

\section{Results}

We computed the posterior probability distributions of the parental masses for both black holes in the binary GW190521 following Eq. \ref{eq:3}. For the prior probability $\pi(M_k)$ we adopted an uninformative, uniform prior, the same that was used by LIGO/Virgo \citep{2020arXiv201014527A}. Integrals involving $M_k$ were approximated as Monte Carlo integrals. For the prior probability $\pi(m_{1},m_{2})$ we adopted the average over the posterior population distribution of the model fit on the GWTC-2 gravitational wave catalog (power law + peak model; \citealt{2021ApJ...913L...7A}). This model takes into account the mass distributions for both black holes, including the correlation between them \citealt{2018ApJ...856..173T}. The prior $\pi(m_{1},m_{2})$ may somewhat underestimate the masses of the parental black holes as dynamical and AGN-assisted formation channels, which are the most likely sites of hierarchical mergers, have top-heavy mass distributions \citep{2019PhRvL.123r1101Y,2021arXiv210704639F}. Other priors are also possible that can give very different mass estimates for GW190521's black holes \citep{2020ApJ...904L..26F,2021ApJ...907L...9N}. 

Results are shown in Fig. \ref{fig:massPDF}. We see that for both black hole masses $M_1$ and $M_2$ within GW190521, one of the parents has a relatively high mass while the lighter parent essentially follows the GWTC-2 distribution. In particular, we find that the heaviest parental black hole has a reconstructed mass $m_{11}=62^{+21}_{-19}$M$_\odot$ (90\% credible level). We find essentially the same $m_{11}$ if we assume a uniform prior distribution for $\pi(m_{1},m_{2})$. This heaviest parent has $78\%$ probability to be in the $50$\,M$_\odot-120$\,M$_\odot$ mass gap, or 43\% if we adopt a more conservative lower limit of $65\,$M$_\odot$. The same probabilities are $86\%$ and $46\%$, respectively, if we assume uniform prior distribution.

\begin{figure*}
\centering
\includegraphics[angle = 0, scale = 0.54]{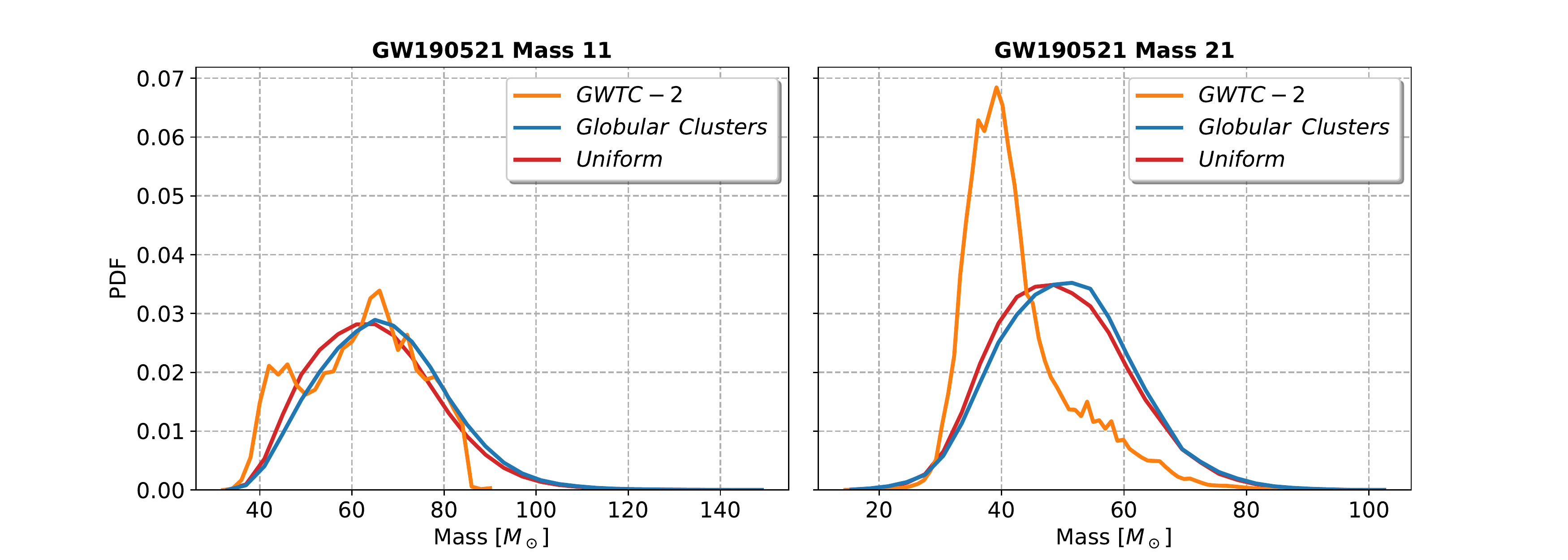}
      \caption{{\bf Mass probability density of the heaviest ancestors of GW190521 for different prior distributions.} We considered the prior distribution from the GWTC-2 gravitational wave catalog (power law + peak model; \citealt{2021ApJ...913L...7A}), an expected distribution for globular clusters based on \cite{2016ApJ...824L..12O}, and a uniform independent prior in $m_1$ and $m_2$.
    \label{fig:priordependene}}
\end{figure*}

To understand the role of the prior distribution in ancestral masses, we considered three different prior distributions for the heaviest parents of both black holes in GW190521. Priors included the GWTC-2 based prior and a uniform prior. We also considered the mass distribution expected in globular clusters \citep{2016ApJ...824L..12O}, where we adopted a probability density $\propto (m_1+m_2)^4(m_1m_2)^{-2.3}$. Results are shown in Fig. \ref{fig:priordependene}.  We find that for the heaviest ancestor, the choice of prior results in negligible difference. Larger difference is found for the heavier parent of $m_2$ in GW190521.

Given the non-negligible probability that at least one of the parents falls in the mass gap, which might indicate that this parent is also the product of a previous merger, we computed the expected masses of GW190521's grandparents as well. The obtained distributions are shown in Fig. \ref{fig:massPDF}. We see that, similarly to parents, the heaviest grandparent has a mass significantly above that of the GWTC-2 distribution, while the lighter grandparent masses follow the GWTC-2 distribution. We further find that, assuming GWTC-2 prior distribution, going back an additional generation, the heaviest great-grandparent has a mass of roughly $30$\,M$_\odot$ and has about same mass uncertainty as the heaviest grandparent.


\section{Conclusion}

We introduced and carried out a Bayesian reconstruction of the ancestral black hole masses of the black holes in the merger GW190521, assuming that GW190521 is the end product of consecutive mergers. We found that, given our prior assumptions, one of the parental black holes has a $78\%$ (43\%) probability of falling in the upper mass gap assuming a lower mass limit of 50\,M$_\odot$ (65\,M$_\odot$), making this parent a plausible candidate for a merger remnant. 

With this possibility in mind we reconstructed the expected mass distribution of grandparent black holes, i.e. the ones two generations before those in GW190521. Remarkably, we found that even after two generations the mass of the heaviest grandparents have limited mass uncertainty, indicating that the reverse engineering of ancestral masses can provide non-trivial information on the ancestors' properties. This information, along with the possible indication that GW190521 might be a 3+ generation merger, might carry indications of the origin of the merger. In particular, galactic nuclei and AGNs appear to be the most likely sites where multiple generations of hierarchical mergers might take place. 

Future detections of even heavier black holes by LIGO/Virgo/KAGRA, and the better theoretical understanding of the pair instability mass gap, could shed further light onto the formation of binary black holes, where the reconstruction of ancestral black hole masses could play an important role.
\newline

\begin{acknowledgments}
The authors would like to thank Zoheyr Doctor, Maya Fishbach, V. Gayathri, Zoltan Haiman, Bence Kocsis and Hiromichi Tagawa for their valuable feedback. O.B. is grateful for the Christopher Schaffer Scholarship Fund  and for the University Scholarship of the University of Florida.  I.B. acknowledges the support of the Alfred P. Sloan Foundation and NSF grants PHY-1911796 and PHY-2110060. This material is based upon work supported by NSF's LIGO Laboratory which is a major facility fully funded by the National Science Foundation. This research
has made use of data obtained from
the Gravitational Wave Open Science Center (\url{https://www.
gw-openscience.org}), a service of LIGO Laboratory, the LIGO Scientific Collaboration and the Virgo Collaboration. LIGO is
funded by the U.S. National Science Foundation. Virgo is
funded by the French Centre National de Recherche Scientifique (CNRS), the Italian Istituto Nazionale della Fisica
Nucleare (INFN) and the Dutch Nikhef, with contributions
by Polish and Hungarian institutes.
\end{acknowledgments}

\bibliography{Refs}
\bibliographystyle{aasjournal}

\end{document}